\documentclass[twocolumn,showpacs,preprintnumbers,amsmath,amssymb]{revtex4}

\usepackage{graphicx}
\usepackage{dcolumn}
\usepackage{bm}
\usepackage[dvips]{color}

\def\mrm{\mathrm}
\def\mbf{\mathbf}

\def\rt{(\mbf{r},t)}
\def\qt{(\mbf{q},t)}

\def\zv{\mbf{0}}
\def\rr{\mbf{r}}
\def\qq{\mbf{q}}

\def\Lsys{L_\mrm{sys}}

\begin{document}


\preprint{APS/123-QED}

\title{
Spatio-temporal scaling for out-of-equilibrium relaxation dynamics 
of an elastic manifold in random media:
crossover between diffusive and glassy regimes
}

\author{Tomoaki Nogawa}
\email{nogawa@statphys.sci.hokudai.ac.jp}

\author{Koji Nemoto}
\email{nemoto@statphys.sci.hokudai.ac.jp}

\affiliation{%
Department of Physics, Hokkaido University,
Sapporo, Hokkaido 060-0810 Japan
}%

\author{Hajime Yoshino}
\email{yoshino@ess.sci.osaka-u.ac.jp}

\affiliation{%
Department of Earth and Space Science, Osaka University,
Toyonaka, Osaka 560-0043 Japan
}%
\date{\today}

\begin{abstract}
We study relaxation dynamics of a three dimensional elastic manifold in random potential from a uniform initial condition by numerically solving the Langevin equation. 
We observe growth of roughness of the system up to larger wavelengths with time. 
We analyze structure factor in detail and find a compact scaling ansatz describing two distinct time regimes and crossover between them. 
We find short time regime corresponding to length scale smaller than the Larkin length $L_c$ is well described by the Larkin model which predicts a power law growth of domain size $L(t)$. 
Longer time behavior exhibits the glassy regime with slower growth of $L(t)$. 
\end{abstract}


\maketitle

\section{Introduction}

Fluctuations around macroscopically condensed states such as
charge density waves (CDW) \cite{Gruner88} and 
flux line lattices in superconductors \cite{Blatter94}, 
often exhibit glassy dynamics due to 
frustration between the elastic restoring forces originated from the
stiffness of the ordered state and random pinning forces brought by impurities.
Thermally activated process dominates the slow relaxation dynamics
of such systems much as spin glasses (see, e. g.,  
\cite{Vincent96, Nordblad98, Bouchaud98, Kawashima04}) 
and super-cooled liquids \cite{Debenedetti01}.

An important basic problem in the studies of the glassy dynamics 
is the isothermal aging, i.e., relaxation at fixed temperature from 
initial states far from equilibrium \cite{Bouchaud98}.  
Aging of elastic manifolds in random media has been studied
theoretically by dynamical mean-field theories 
\cite{Cugliandolo96}
and numerical simulations 
\cite{Yoshino96, Yoshino98, Kolton05, Kolton06, Bustingorry06, Bustingorry07, Yoshino0x}.
Presumably there exists a dynamical length scale $L(t)$ 
which grows with time $t$ such that the system is equilibrated 
on the wavelengths smaller than $L(t)$ \cite{Paul04,Rieger05,Yoshino0x}. 
In other word $L(t)$ is the size of local equilibrium domain. 
Roles of $L(t)$ have been examined extensively in the context of
aging of spin-glasses 
(see for instance, Ref. \cite{Kisker96, Komori99, Huse85}).
While $L(t)$ grows algebraically without the random pinning forces,
the frustration drastically slows it down. Typically one expects that
the growth law becomes logarithmic 
due to the energy barriers which grow with the length scale 
\cite{Villain84, Huse85, HuseFisher87, FisherHuse91, Mikheev95, 
Kolton05, Yoshino0x}.
The purpose of the present paper is to analyze aging
of an elastic manifold in random media in terms of $L(t)$
and investigate crossovers between the two characteristic regimes: 
so called the Larkin regime and glassy regime.

A standard theoretical model to study the above mentioned CDW-{\it like} 
systems is the elastic manifold model in random potential, 
e.g. the Fukuyama-Lee-Rice model \cite{Fukuyama78, Lee79}, 
given by the following Hamiltonian 
\begin{equation}
H=\int d\mbf{r} \left[ \frac{1}{2} \kappa 
\left|\nabla \theta(\mbf{r},t) \right| ^2
- h(\rr) \cos\left( \theta ( \rr, t) - \beta(\rr) \right) \right]. 
\label{eq:rfxy_hamiltonian}
\end{equation}
Physically the scalar field $\theta(\rr)$ at position $\rr$ in the space 
are understood as the local fluctuation of the phase part of order parameter 
of the condensate, such as the CDW state. 
The first term with elastic constant $\kappa$ 
indicates the elastic deformation energy 
which is minimized when $\theta(\rr,t)$ is spatially uniform 
in the absence of the second term, the random field energy. 
This sinusoidal random potential is a periodic function 
with respect to $\theta(\rr,t)$ reflecting the underlying periodicity
of the condensate. 
Both amplitude $h(\rr)$ and phase $\beta(\rr)$ of the random-field 
are quenched random variables with short-ranged spatial correlations. 
Hereafter $\langle \ldots \rangle$ means a thermal average 
and $\overline{\cdots}$ means an average over the quenched randomness 
(samples).

Let us recall here some basic {\it static} properties of the system
which is understood better than the dynamical properties of our interests.
It is believed that physical properties 
of this kind of systems,
such as the roughness characterized by
$B(\rr)=\overline{ \langle ( \theta(\rr) - \theta(\zv) )^2 \rangle}$,  
are different on three distinct length scales \cite{Giamarchi95}.
First, in very short length regime, 
perturbative analysis of the effects of disorder predict 
algebraic growth of the roughness with distance $r$:
$B(\rr) \propto r^{2\zeta}$ with some roughness exponent
$\zeta$. Below four dimensions 
it is known that  $\zeta=(4-d)/2$ \cite{Larkin70, Imry75, Fukuyama78}.
Then the perturbative regime, which we call as the {\it Larkin regime} 
in the following,  must be terminated
at the so-called Larkin length $L_c$ 
over which the effect of randomness overcomes the elasticity.
Then the so called {\it random manifold regime} begins 
\cite{Fisher85,Fisher86,Feigelman89,Bouchaud-Mezard-Yedidia91} 
where many metastable states exist and 
the roughness of the system is characterized 
by a nontrivial roughness exponent $\zeta_\mrm{rm}$. 
In much larger length scales, amplitude of $\theta$ eventually grows 
beyond the period of the random potential. 
If the periodicity is relevant, 
the system cannot gain more benefit of the potential energy 
at a cost of elastic energy. 
Then the last regime called the Bragg glass regime begins.
In three dimensions large wave length fluctuation is highly suppressed 
and $B(\rr)$ is no longer expressed by algebraic functions $r^{2\zeta}$
but by a certain logarithmic function of the distance $r$
\cite{Nattermann90, Korshunov93, Giamarchi95}.  
Then it is said that the system is in the Bragg glass phase 
where the system has a quasi-long-range order (QLRO). 
In the case of single harmonic potential as the present model 
Eq.~(\ref{eq:rfxy_hamiltonian}), 
the end of the Larkin regime and the beginning 
of the Bragg glass regime coincide, i.e., 
the transient random manifold regime does not exist 
\cite{Giamarchi95,Giamarchi98}. 
This is because the potential has only single characteristic scale, 
that is a period $2\pi$,  
and does not have another smaller scale 
which yields the upper bound of the Larkin regime 
such as short ranged correlation length 
of potential along $\theta$-direction.

In this paper, we study the 
out-of-equilibrium relaxation dynamics 
of the elastic system in the periodic random potential, 
Eq.~(\ref{eq:rfxy_hamiltonian}). 
This system shows different types of dynamics 
at different time scales. 
We show that each of these is related to 
equilibrium spatial property in the corresponding length scale.
Although we discuss in particular the case of a three dimensional system 
we may comment on systems in general $d$-dimensions.


In the next section, 
we review the dynamics in the Larkin regime, 
which can be examined analytically. 
In the section III, numerical analysis 
of the structure factor is shown. 
In the section IV, we propose a scaling law which 
describes the crossover from the Larkin regime to the 
glassy regime. 
In the final section, we present conclusions and remarks.

\section{Power-low domain growth in Larkin regime}
\label{sec:larkin}


When phase fluctuation is very small, $\theta(\rr,t) \ll 1$, 
the Hamiltonian Eq.~(\ref{eq:rfxy_hamiltonian}) is reduced to 
the so-called Larkin model \cite{Larkin70}, 
which is exactly solvable. 
When the second term in the r.~h.~s. of Eq.~(\ref{eq:rfxy_hamiltonian}) is 
expanded by $\theta$ up to the linear order, 
the disorder effect is represented by 
the quenched random force $\eta(\mbf{r})=-h(\rr) \sin \beta(\rr)$, 
which is supposed to be a random Gaussian number satisfying  
\begin{equation}
\overline{\eta(\mbf{r})}=0, \quad
\overline{\eta(\mbf{r}) \eta(\mbf{r}')}=\Delta_h \delta(\mbf{r-r'}), 
\end{equation}
with a finite variance $\Delta_h$.
The overdamped Langevin equation is written as 
\begin{eqnarray}
\gamma \frac{d}{dt} \theta\rt 
&=& -\frac{\delta H}{ \delta \theta\rt} + \zeta\rt
\nonumber \\
&=& \kappa \nabla^2 \theta\rt + \eta(\mbf{r})+ \zeta\rt, 
\label{eq:larkin-model}
\end{eqnarray}
where $\gamma$ is the friction coefficient 
and $\zeta\rt$ is a Gaussian white thermal noise satisfying  
\begin{equation}
\langle \zeta\rt \rangle=0, \quad 
\langle \zeta\rt \zeta(\mbf{r}',t') \rangle
= 2 \gamma T \delta(\mbf{r-r'})\delta(t-t'), 
\end{equation}
$T$ being the temperature of the heat bath. 
The average $\langle \cdots \rangle$ is taken 
over independent noise realizations. 

The formal solution with the uniform initial condition, 
$\theta(\mbf{r},0)=0$ for all $\rr$, is expressed as 
\begin{equation}
\theta\qt = \frac{1}{\gamma} \int_0^t dt' e^{-\kappa q^2 (t-t')/\gamma} 
\left[ \eta(\mbf{q}) + \zeta(\qq,t') \right], 
\end{equation}
where the functions of $\qq$ are the Fourier transformations 
of the corresponding functions of $\rr$. 
The structure factor, i.e., Fourier transform 
of the scalar correlation function is obtained as 
\begin{eqnarray}
B\qt  \equiv  
\int d \rr \overline{ \langle \theta(\mbf{0},t) \theta\rt \rangle } 
e^{i \mbf{q \cdot r}}
= \overline{ \langle  |\theta\qt|^2 \rangle }
\nonumber \\
\qquad =  T \frac{ 1- e^{- 2 q^2 L(t)^2} }{\kappa q^2}
+ \Delta_h \left[ \frac{ 1- e^{- q^2 L(t)^2} }{ \kappa q^2 } \right]^2, 
\label{eq:sq_el}
\end{eqnarray}
where 
\begin{equation}
L(t) \equiv \sqrt{\kappa t / \gamma}. 
\label{eq:power_growth}
\end{equation}
Equation (\ref{eq:sq_el}) means that the two fluctuations 
owing to temperature and randomness are decoupled in the Larkin model. 
But the growth of the two are characterized 
by a single time-dependent length $L(t)$ 
such that the system is equilibrated over wavelength 
shorter than $L(t)$, i.e., 
$B(\qq,t) \sim B(\qq,t\rightarrow \infty) 
= T/\kappa q^2 + \Delta_h/\kappa^2q^4$ 
for $q L(t) \gg 1$.

From Eq.~(\ref{eq:sq_el}), 
we can evaluate the amplitude of phase fluctuation as 
\begin{eqnarray}
\sigma(t) &\equiv&
\overline{\langle \theta(\rr,t)^2 \rangle} 
= \int d\qq B(\qq,t) 
\\
&\approx&  C_0 - \frac{C_1}{d-2} \frac{T}{\kappa} L(t)^{-(d-2)} 
+ \frac{C_2}{4-d} \frac{\Delta_h}{\kappa^2} L(t)^{4-d} ,
\label{eq:theta_squared}
\end{eqnarray}
where $C_1$ and $C_2$ are positive constants.
The third term diverges in equilibrium 
($t \rightarrow \infty$, $L(t)\rightarrow \infty$) 
below four dimensions. 
By writing $\sigma(t) \propto L(t)^{2\zeta}$
one can read off the the roughness exponent of the Larkin model
as $\zeta_\mrm{Larkin}=(4-d)/2$.
In the long length (time) scale,  
$L(t) \gg L_T \equiv \sqrt{\kappa T/ \Delta_h}$ , 
the third term due to the quenched randomness is dominant
and the second term due to the thermal fluctuation term can be ignored.

Note that the growth law of $L(t)$ given by Eq.~(\ref{eq:power_growth})
is the same as in the absence of the random potential, i. e., 
diffusive dynamics with dynamical exponent $z=2$.
It means that the quenched random potential does not bring pinning effects 
at the level of its linear approximation. It is easy to see that
this is the case at any higher levels of perturbative treatments 
of the random potential.

\section{Glassy dynamics in Nonlinear Potential}

The linear approximation adopted in the previous section 
breaks down for large $\sigma(t)$, 
which occurs when $L(t)$ becomes as large 
as the so-called Larkin length, $L_c$;
\begin{eqnarray}
L_c(\Delta_h) = 
\left( \frac{C_2}{c_L^2  \kappa^2 } \Delta_h \right) ^{-1/(4-d)}
.
\label{eq:lc}
\end{eqnarray}
This length is derived from the threshold condition;  
$\sigma(t) = ( 2\pi c_L )^2$, 
where $c_L$ is a constant, similar to the Lindemann's constant
\cite{Lindemann10}, 
indicating the border below which the nonlinearity of random potential 
can be ignored. 
The corresponding time scale is 
\begin{equation}
t_c(\Delta_h) = L_c(\Delta_h)^2 \gamma/\kappa 
\propto \Delta_h^{-2/(4-d)}.
\label{eq:tc}
\end{equation}

Beyond the Larkin length 
nonlinearity of the random potential yields many metastable states. 
The potential barriers between them 
will significantly slow down the growth of $\sigma(t)$. 
Hereafter we call this nonpurturbative regime just as 
{\it the glassy regime} 
which may corresponds to the random manifold or Bragg glass regimes.
Now we study the growth of roughness in the glassy regime 
by numerical simulations.

\subsection{Simulations}
In practice 
we consider the lattice version of Eq.~(\ref{eq:larkin-model}). 
The equation of motion for the phase at the lattice point $\rr_i$ is 
\begin{eqnarray}
\Gamma \frac{d}{dt} \theta_i(t) 
&=& - J \sum_{j\in \mrm{n.n.}}  \left[ \theta_i(t) - \theta_j(t) \right]
\nonumber \\
&& \quad - h_i \sin \left( \theta_i(t) - \beta_i \right) 
+ \zeta_i(t), 
\label{eq:eom2}
\end{eqnarray}
where
\begin{eqnarray}
\langle \zeta_i(t) \zeta_j(t') \rangle 
&=& 2 \Gamma T \delta_{ij} \delta(t-t').
\end{eqnarray}
Hereafter we set the coupling constant $J$ 
and the friction coefficient $\Gamma$ to unity. 

Here let us explain some details of the numerical simulations. 
Phase variables $\theta_i(t)$'s are put on 
the cubic lattice with size $N={\Lsys}^3=128^3$ 
and periodic boundary conditions are imposed in all directions. 
The phase $\beta_i$'s are introduced as independent uniform random numbers 
between 0 and $2\pi$. 
On the other hand 
the strength of the random field $h_i$ is set 
to a uniform value $h$ for all sites
\footnote{
We consider the distribution of the amplitude $h_i$ on each site 
does not change the semi-quantitative property in the weak pinning regime 
because the system feels averaged random potential 
over the region where phase is almost uniform. 
We checked it by preliminary simulations. 
}, so that 
\begin{equation}
\Delta_h = h^2/2.
\end{equation}

We investigate the relaxation dynamics at various values
of the random field $h=1.0-5.0$ and temperatures
$T=0.0, 0.5, 1.0$ and $1.5$
\footnote{
These temperatures are lower 
than the ferromagnetic transition temperature, $\sim 2.20$, 
of the pure XY model on the cubic lattice, 
whose spin-wave-approximation is the present model. 
}.
We numerically solve the Eq.~(\ref{eq:eom2}) 
by the second order stochastic Runge-Kutta method \cite{Honeycutt92}. 
In the initial state, 
$\theta_i$ is set to $0$ for all $i$. 
Physical quantities are averaged over $128$ runs at least,
each of which has independent realizations 
of random phase $\beta_i$ and thermal noise $\zeta_{i}(t)$. 
 
We define the lattice version of the structure factor 
Eq.~(\ref{eq:sq_el}) as
\begin{eqnarray}
B(\qq,t) = N \overline{\langle 
\left| \theta(\qq,t) \right| ^2 
\rangle }
\label{eq:bqt-descrete}
\end{eqnarray}
with
$\theta(\qq,t) = N^{-1} \sum_j \theta_j(t) e^{i \qq \cdot \rr_j}$ 
\footnote{
More precisely we calculated 
$B(q_0,t)=[B(\qq_0=(q_0,0,0),t)
+B(\qq_0=(0,q_0,0),t)+B(\qq_0=(0,0,q_0),t)]/3$ 
where $q_0 = 2\pi n /\Lsys$  for $n = 0,1,2,\cdots,L/2$.
Hereafter we use 
$q^2 = 2(1-\cos q_0) = q_0^2 + O(q_0^4)$ instead of $q_0^2$. 
By using this $q$, the same formula with that of continuum model, 
such as Eq.~(\ref{eq:sq_el}), can be used 
for the Larkin model on lattice. }.

\subsection{Results}

\begin{figure}
\begin{center}
\includegraphics[trim=50 250 130 -240,scale=0.4,clip]{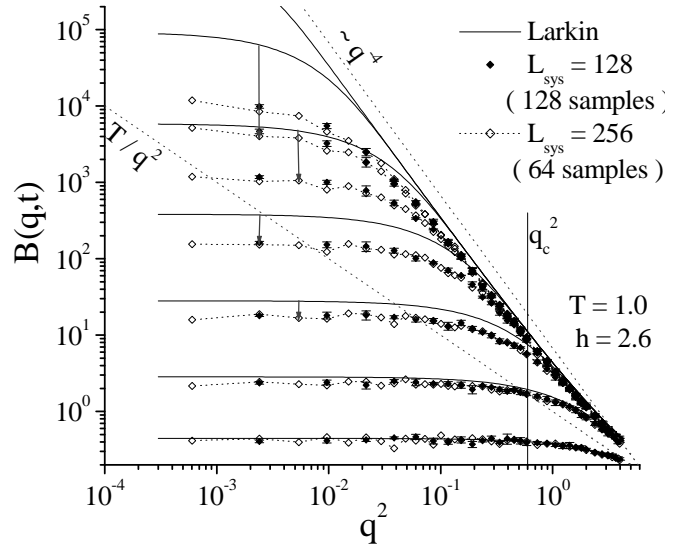}
\end{center}
\vspace{-5mm}
\caption{\label{fig:bqt-q2}
Snapshots of the profile of the structure factor $B(q,t)$.
The time changes uniformly in logarithmic scale 
as $t/0.34 = 4^0, 4^1,\cdots, 4^6$ from the bottom to the top. 
The symbols with error bars are data for $\Lsys=128$ 
and dotted lines are for $\Lsys=256$. 
The arrows connect the values in the Larkin model (bold curves)
given by Eq.~(\ref{eq:sq_el})
and numerical data obtained in the sinusoidal model at the same time $t$. 
The vertical line shows $q^2 = t_c^{-1}$ [47].
}
\end{figure}

Figure \ref{fig:bqt-q2} shows some 
examples of the profile of $B(q,t)$.
Components for all $q$ are zero at $t=0$ 
and the structure factor grow with time $t$. 
It can be seen that at larger $q$ the amplitude 
saturates to $t$-independent but $q$-dependent value 
while at smaller $q$ the amplitude 
remains $t$-dependent but $q$-independent.
The above observation suggests that there is indeed a dynamical 
length scale $L(t)$ which grows with time $t$ such that components 
satisfying $qL(t) \gg 1$ become equilibrated: the system has become
rough on short wavelengths but remains flat at larger wavelengths.

A simplest scaling which connects the dynamical regime
$q L(t) \ll 1$ and the static regime $q L(t) \gg 1$ may be 
\cite{Yoshino98, Schehr-Doussal05, Kolton05, Kolton06, Yoshino0x}
\begin{equation}
B(q,t)= q^{-(d+2\zeta)}\tilde{B}(qL(t))
\label{eq:simple-scaling}
\end{equation}
where $\zeta$ is the roughness exponent and the scaling function
behaves as $\tilde{B}(x) \sim \rm{const.}$ for $x \gg 1$ and
$\tilde{B}(x) \sim x^{d+2\zeta}$. 
By integrating over $q$ 
one obtains the corresponding scaling form for 
\begin{equation}
\sigma(t) \equiv \sum_{\qq \ne \mbf{0} } B(\qq,t) 
= \overline{\langle \theta(\rr,t)^2 \rangle}
\label{eq:sigma_lattice}
\end{equation}
as
\begin{equation}
\sigma(t) \propto L(t)^{2\zeta}.
\label{eq:simple-sigma-scaling}
\end{equation}
(Note that the above scaling holds only if 
the system size $\Lsys$ is sufficiently larger than $L(t)$ 
for a given time $t$.)
Indeed one can find easily 
that the Larkin model discussed in section \ref{sec:larkin}
satisfies these scalings exactly at $T=0$.
However, the real behavior will be 
more complicated even in the Larkin regime because roughness originates 
not only from the quenched random field but also from the thermal noise 
at finite temperatures. Furthermore, there will be a crossover from the Larkin 
regime to the glassy regime at $L_{c}$.
In section \ref{sec:scaling} we perform a more elaborate scaling 
analysis taking into account these complications.

The analytic solution of the Larkin model 
is also plotted in Fig.~\ref{fig:bqt-q2} for comparison. 
In very short time the structure factors of the two models coincide. 
As time goes by, it becomes apparent that 
$B(q,t)$ for the sinusoidal potential model 
Eq.~(\ref{eq:eom2}) does not grow as fast as that of the Larkin 
model Eq.~(\ref{eq:larkin-model}).  
As shown later, the time scale beyond which 
the equivalence breaks down is $t_{c}$ given by Eq.~(\ref{eq:tc}). 
Furthermore by a closer inspection it appears that
the envelope function $B(q,t \to \infty)$, 
i.e., the equilibrium structure factor, changes 
from that of the Larkin model. 
The structure factor for small $q$ parts seems to be slightly different 
from $q^{-4}$ of the Larkin model.
We regard these changes as the crossover from the Larkin regime 
to the glassy regime, which  
we analyze more carefully
in the section \ref{sec:scaling}. 

\begin{figure}
\begin{center}
\includegraphics[trim=50 250 100 -200,scale=0.38,clip]{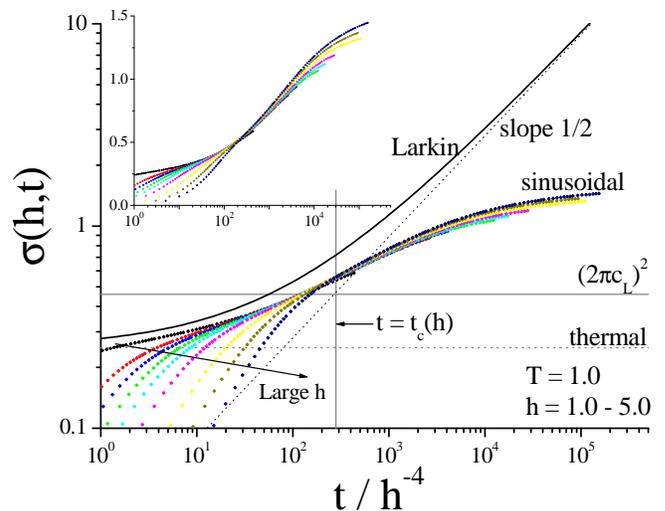}
\end{center}
\vspace{-5mm}
\caption{\label{fig:sigma-t}
Time evolution of $\sigma(t)$. 
The inset is a semilogarithmic plot of the same data. 
The solid line curves show the results of the Larkin model 
for $h \ll T$. 
The horizontal dotted line indicates the equilibrium 
value $\sigma(t=\infty)$ for $h=0$. 
The lines noted with $2\pi c_L$ and $t_c(h)$ 
indicate the crossover (see detail in [47]).
}
\end{figure}

Figure \ref{fig:sigma-t} shows the time evolution of $\sigma(t)$. 
In the Larkin regime $L(t) \ll L_{c}$ or $t \ll t_{c}$, 
we expect $\sigma(t) \propto L(t)^{2\zeta} \sim t^{2\zeta/z}$ 
(See Eq.~(\ref{eq:simple-sigma-scaling}))
with $\zeta=1/2$ and $z=2$. 
However, the data deviate from this behavior for $t>t_c$. 
The crossover time
increases as the strength of the random field $h$ decreases.
We consider this reflects crossover from the Larkin regime to
the glassy regime. 
Indeed by simply scaling $t$ by the anticipated crossover 
time $t_{c} \propto h^{-4}$ given in Eq.~(\ref{eq:tc}), 
data of $\sigma(t)$ collapse onto a universal function 
for sufficiently large $t$. 
The growth of $\sigma(t)$ for $t>t_c$ is very slow presumably due
to activated glassy dynamics.

\section{Scaling of Structure factor}
\label{sec:scaling}


As observed in the previous section, the crossover from the
Larkin regime to the glassy regime
can be described by a simple scaling in which the time
$t$ is scaled by the crossover time $t_c(\Delta_h)$ given by Eq.~(\ref{eq:tc})
corresponding to the Larkin length $L_c(\Delta_h)$.
Now we analyze the spatio-temporal scaling law 
of the structure factor $B(q,t)$ itself, which provides
us more detailed information than the integrated one $\sigma$ 
in Eq.~(\ref{eq:sigma_lattice}).
The basic idea is expressed by Eq.~(\ref{eq:simple-scaling})
which connects the dynamic $q L(t) \ll 1$ and static $q L(t) \gg 1$ regimes. 
However
we need to take into account complications due to the 
roughness of different origins: 
i) thermal roughness with the roughness
exponent $\zeta_{\rm thermal}=(2-d)/2$,  
ii) roughness due to the random field in the Larkin regime 
with $\zeta_{\rm Larkin}=(4-d)/2$ 
and iii) roughness in the glassy regime 
which has different $q$-dependence with ii).

\subsection{Scaling ansatz}

We propose the following scaling ansatz. Most importantly
the crossover from the Larkin to the glassy regime
is taken into account by scaling the length (or the wave number) 
by the Larkin length $L_c$ given by Eq.~(\ref{eq:lc}) and the time 
by the corresponding time scale $t_{c}$ given by Eq.~(\ref{eq:tc}). 
We propose that the structure factor takes the following form, 
\begin{eqnarray}
B(q,t) &=& T L_c^2 \frac{1-e^{-2Y(qL_c)X(t/L_c^2)}}{Y(qL_c)}
\nonumber \\
&+& \Delta_h L_c^4 \left[ \frac{1-e^{-Y(qL_c)X(t/L_c^2)}}{Y(qL_c)} \right]^2. 
\label{eq:scaling1}
\end{eqnarray}
This is an extended version of Eq.~(\ref{eq:sq_el}): 
the first and second terms describe the fluctuation due to thermal noise 
\footnote{
Strictly speaking the first term in Eq.~(\ref{eq:scaling1}), 
the structure factor of the purely thermal origin, 
does not need to be scaled in the same as the second term. 
(One can argue that the purely thermal roughness exists only  in the range
$qL_{c} \gg 1$.)
This is an artifact of our scaling form, which is chosen for simplicity
but it does not make significant changes
on the analysis at small $q$ regimes of our interest
where the second term is dominant. 
}
and quenched randomness, respectively. 
The dynamical and static regimes described in the simplest scaling
Eq.~(\ref{eq:simple-scaling}) correspond to $XY \ll 1$ and $XY \gg 1$
respectively: in the dynamical regime $XY \ll1$
the structure factor is $t$-dependent but $q$-independent while
in the static regime  $XY \gg 1$, it becomes $t$-independent but $q$-dependent.

We suppose that the scaling functions $X$ and $Y$ take 
the following asymptotic forms in the Larkin regime,
\begin{eqnarray}
X(\tilde{t}) = \Bigg\{
\begin{array}{ccc}
\tilde{t} & \mrm{for} & \tilde{t} \ll 1 \\
X_g(\tilde{t})
& \mrm{for} & \tilde{t} \gg 1, 
\end{array}
\\
Y(\tilde{q}) = \Bigg\{
\begin{array}{ccc}
\tilde{q}^2 & \mrm{for} & \tilde{q} \gg 1\\
\mrm{cst.} \times  
\tilde{q}^{d/2+\zeta_g} 
& \mrm{for} & \tilde{q} \ll 1.
\end{array}
\end{eqnarray}
The scaling function $Y(\tilde{q})$ with
$\tilde{q}=qL_{c}$ describes the equilibrium structure factor: 
$Y(\tilde{q})^{-2}=\tilde{q}^{-4}$ 
in  the Larkin regime $\tilde{q} \gg 1$ and 
$\tilde{q}^{-(d+2\zeta_g)}$ in the glassy regime 
$\tilde{q} \ll 1$. 
The exponent $\zeta_g$ is an unknown roughness exponent 
in the glassy regime which will be smaller 
than $\zeta_{\rm Larkin}$. 
Particularly $\zeta_g$ will be zero if 
the system has quasi long range order. 
On the other hand,
the scaling function $X(\tilde{t})$ with $\tilde{t}=t/t_{c}$
describes the growth law of $L(t)$
in the Larkin regime $\tilde{t} \ll 1$ and the glassy regime 
$\tilde{t} \gg 1$. 
More precisely, the dynamical length $L(t)$ can be estimated 
by solving 
\begin{equation}
Y(L(t)^{-1} L_c)X(t/L_c^2) =1.
\end{equation}
The function $X_g(\tilde{t})$ 
is an unknown increasing function of the scaled time $\tilde{t}$ 
in the glassy regime
which will be slower than any algebraic functions 
due to the anticipated activated dynamics \cite{Kolton05, Yoshino0x}.

For $\tilde{t} \ll1$ and $\tilde{q} \gg1$, the above scaling
reproduces Eq.~(\ref{eq:sq_el}) in the Larkin regime. 
However it turns out that certain vertex corrections are 
needed in the coefficients as 
\begin{equation}
T \rightarrow T ( 1 + \alpha_{\scriptscriptstyle T\Delta} \Delta_h 
+ \cdots)
\label{eq-T-correction}
\end{equation}
and 
\begin{equation}
\Delta_h \rightarrow \Delta_h( 1 - \alpha_{\scriptscriptstyle \Delta T} T 
+ \cdots)
\label{eq-h-correction}
\end{equation}
in analyzing the raw data.
These coefficients appear when performing perturbation expansion 
of the random potential 
beyond the linear approximation in section \ref{sec:larkin}. 
In the following analysis 
we treated the first order correction terms only 
and regarded them as fitting parameters.

\begin{figure}
\begin{center}
\includegraphics[trim=20 240 70 -220,scale=0.35,clip]{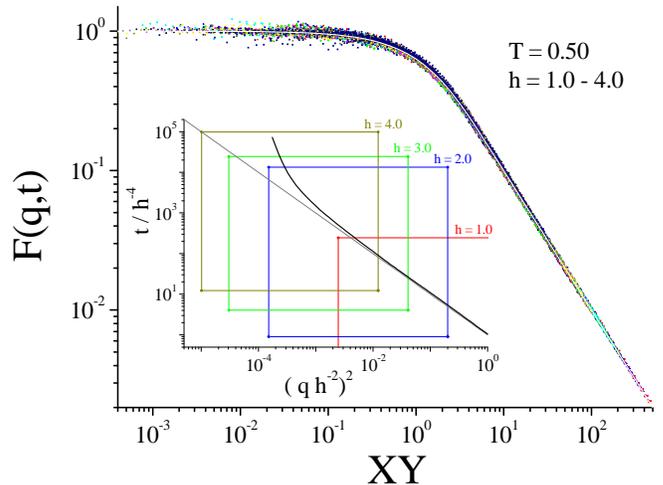}
\end{center}
\vspace{-5mm}
\caption{\label{fig:db}
Result of the scaling at $T=0.5$. 
$F(q,t)$ is plotted as a function of $X(t/t_c)Y(q L_c)$ 
for $h$=1.0, 1.6, 2.0, 2.5, 2.7, 3.0, 3.2, 3.5 and 4.0. 
The white curve indicates $(1-e^{-XY})/XY$.
Here the coefficients are corrected with 
$\alpha_{\scriptscriptstyle T \Delta}=0.027$ 
as in Eq.~(\ref{eq-T-correction})
and $\alpha_{\scriptscriptstyle \Delta T}=0.22$ 
as in Eq.~(\ref{eq-h-correction}).  
The rectangles in the inset show the data ranges of our simulations 
in $t/t_c$ and $q^2 L_c^2$ for $h$=1.0, 2.0, 3.0 and 4.0. 
The bold curve in the inset indicates $X(t/t_c)Y(qL_c)=1$. 
The space-time region above this curve is equilibrated. 
}
\end{figure}

\subsection{Numerical analysis}

Now let us examine the validity of the scaling law presented above 
using our numerical data.
Scaling functions $X(\tilde{t})$ and $Y(\tilde{q})$ are determined 
by least square fitting. 

First we perform fitting by fixing the temperature $T$. 
For example, Fig.~\ref{fig:db} shows the result of the scaling 
for $T=0.5$. We plotted $F(q,t)$ as a function of $XY$ where
\begin{equation}
F(q,t)= \frac{1}{X} \sqrt{ 
\frac{B(q,t)}{L_c^4 \Delta_h} 
- \frac{T}{L_c^2 \Delta_h} \frac{1 - e^{-YX}}{Y}  ,
} 
\nonumber
\label{eq:scaling2}
\end{equation}
which leads $(1-e^{-XY})/XY$ 
if the scaling law in Eq.~(\ref{eq:scaling1}) is valid. 
The figure shows nice collapsing of data 
using proper scaling functions, $X$ and $Y$.

\begin{figure}
\begin{center}
\includegraphics[trim=0 230 60 -200,scale=0.30,clip]{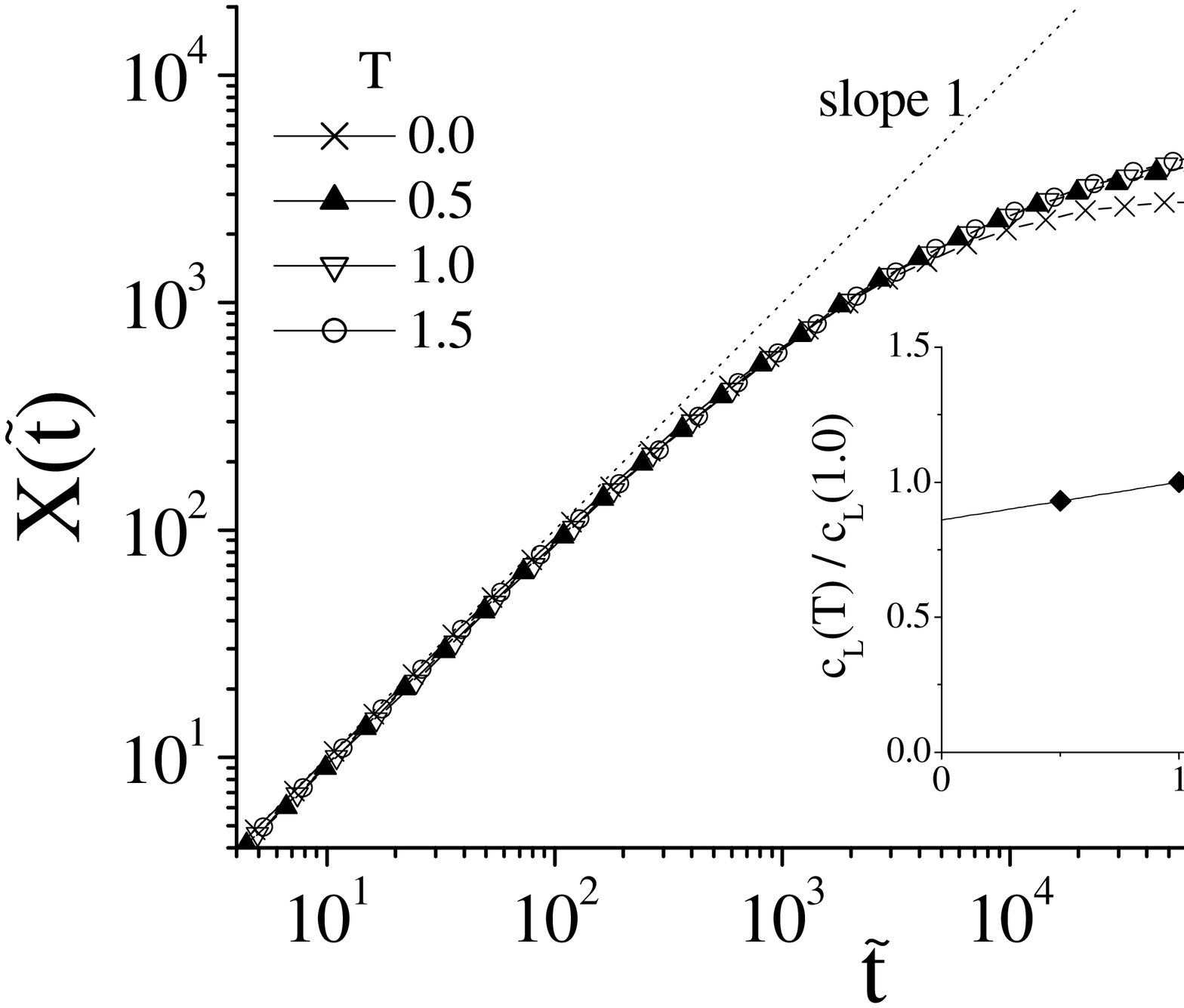}
\includegraphics[trim=0 230 60 -200,scale=0.30,clip]{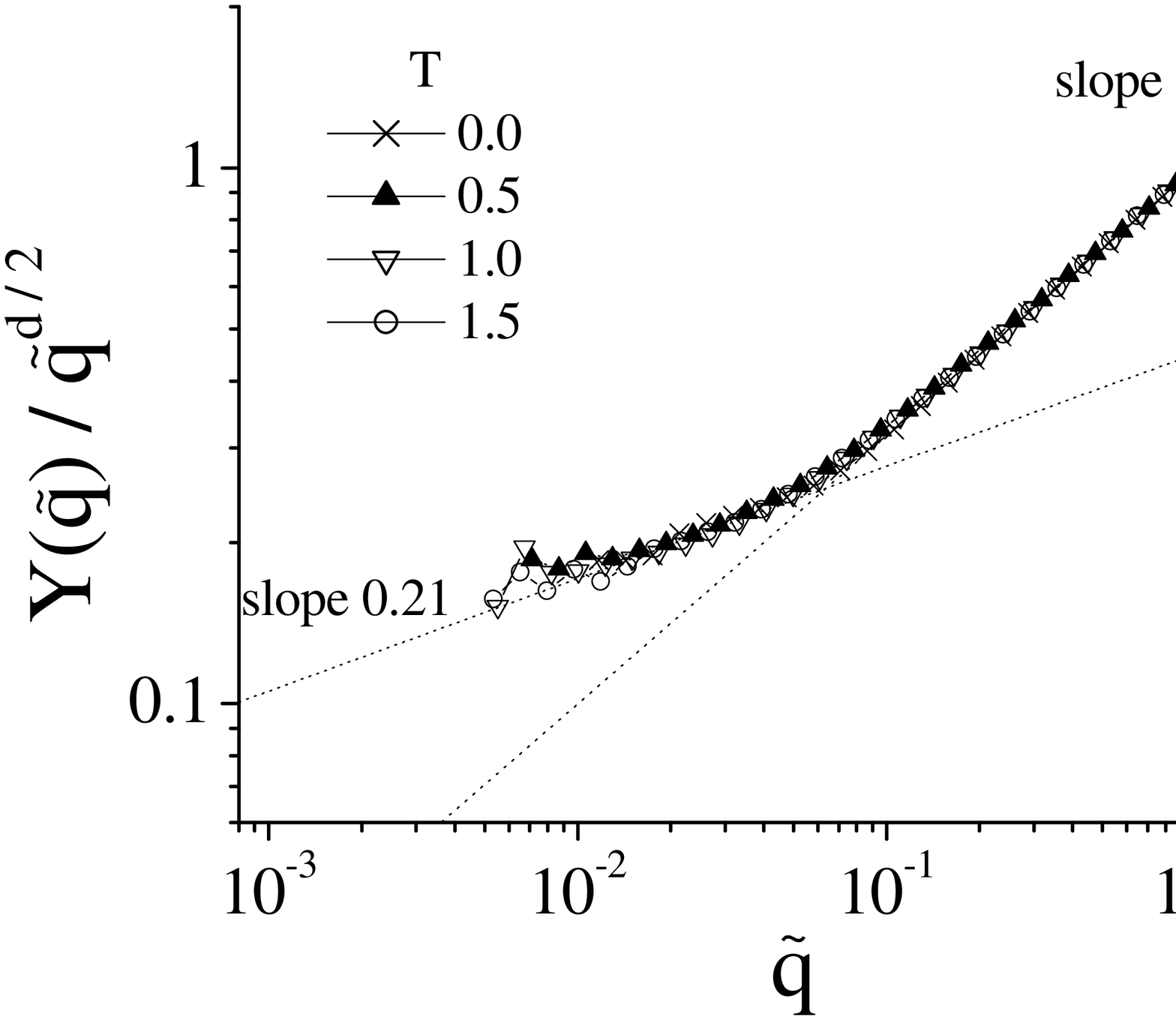}
\end{center}
\vspace{-5mm}
\caption{\label{fig:X-x}
Scaling function $X(\tilde{t})$ and $Y(\tilde{q})$ 
for several fixed temperatures. 
The inset in the top graph indicates the correction factor $c_L(T)$.
Here we use $c_L(T=0)$ by extrapolating from finite temperature data. 
More precisely speaking the horizontal axes in this plot, 
$\tilde{t}$ and $\tilde{q}$, are scaled value using 
$L_c(\Delta_h,T) = [h c_L(1.0) / c_L(T)] ^{-2}$
instead of that in Eq.~(\ref{eq:lct}).
}
\end{figure}

Next let us bring together data at different temperatures.
To this end we note that the Larkin length can weakly depends
on temperature,
\begin{equation}
L_c(\Delta_h,T) = [\Delta_h / c_{L}(T)^2] ^{-1/(4-d)}.
\label{eq:lct}
\end{equation}
The constant $c_{L}(T)$ will be larger for higher temperatures
because thermal fluctuations will mask the quenched random potential
over short length scales. In practice we treated $c_{L}(T)$ 
as a fitting parameter.
Then as shown in Fig.~\ref{fig:X-x}, 
scaling function $Y$ for different temperatures
can be laid on a universal curve independent of temperatures.
The resultant correction factor $c_{L}(T)$ is shown 
in the inset of Fig.~\ref{fig:X-x}.

The scaling function $X(\tilde{t})$ exhibits the crossover 
from the power-law domain growth to the glassy dynamics. 
Looking Fig.~\ref{fig:X-x} carefully, 
the scaling function $X$ is not universal, 
which is more apparent in lower temperature. 
Particularly the relaxation stops on the way at zero temperature. 
This indicates that the relaxation in the long time regime 
is thermally activated process.  
The system cannot escape from a metastable state without thermal assistance.

The scaling function $Y^{-2}$ represents 
the structure factor in the equilibrium state ($X \rightarrow \infty$). 
From the present scaling we obtain its shape even for the 
small wave numbers $\tilde{q} \ll 1 $ where $B(q,t)$ is still 
far from equilibrium. 
While the growth law $X(t)$ turned out to depend on the temperature $T$, 
we find that $Y(t)$ is essentially independent of the temperature $T$.
This means that 
spatial correlation function in equilibrium 
has universal form independent of both $T$ and $\Delta_h$. 
The long wave length behavior of $Y$ seems to obey 
a power law; $Y(\tilde{q})\sim \tilde{q}^{-(d/2+\zeta)}$.  
The roughness exponent $\zeta$ is smaller than that in the Larkin regime 
and the value is consistent with 
the one for the random manifolds $\zeta_\mrm{rm}\approx 0.21$
\footnote{
The roughness exponent for the $d$-dimensional system with 
$N$-components deformation field is written as 
$\zeta_\mrm{rm}(d,N) = (4-d)/[4 + \nu(d,N) N]$ \cite{Nattermann00}. 
The exponent $\nu(3,1)$ corresponding to the present system 
has not known but expected to be between 
$\nu(1,1)=1/2$ \cite{Huse85} and $\nu(4,1)=0.80$ \cite{Balents93}.
Substituting these $\nu$'s and $d=3$ to the above formula 
instead of $\nu(3,1)$ leads close values of $\zeta_\mrm{rm}, 0.22$ 
and $0.21$, respectively. 
}. 
However it is more natural to expect 
that this agreement is a transient behavior 
that $\zeta$ is approaching to zero 
because the present single harmonic model 
is considered to take single crossover 
to the Bragg glass regime \cite{Giamarchi95,Giamarchi98}. 



\section{Summary and Discussions}

In this paper the relaxation dynamics of 
the three dimensional elastic manifold in random potential  
has been studied. 
We especially focused on the crossover 
between the Larkin regime and the glassy regime, 
i.e.,  power-law domain growth and thermally activated relaxation. 
We proposed a new scaling method for the dynamical structure factor 
which encodes dynamical growth of 
the roughness of different origins 
and successfully applied it to explain the crossover between the
Larkin and glassy regime. 
At a given temperature the structure factor out-of-equilibrium 
can be scaled by using the dynamical length $L(t)$ 
and the Larkin length $L_c(\Delta_h)$. 
Quite interestingly our analysis yields the structure factor in equilibrium, 
which is hard to observe in equilibrium simulations. 
The temperature dependence can be also taken care
only by introducing a correction factor $c_{L}(T)$ for the Larkin length. 
It turns out that the scaling function $Y(\tilde{q})$ 
are universal and independent of the temperatures.

Although we analyzed the model with random potential 
that is periodic with respect to the deformation field, 
indication of QLRO was hardly observed. 
This will appear when the amplitude of phase fluctuation 
becomes greater than the period of the random potential, 
$\sigma(t) \gg (2\pi)^2$. 
The crossover region between the Larkin and Bragg glass regimes, 
however, persists for quite long time 
and has special importance in the dynamical aspect. 
This is because glassy behavior becomes serious 
at the early stage of the crossover. 
From the obtained scaling function $Y(\tilde{q})$, 
we can roughly estimate the end of the Larkin regimes 
as $c_L=\sqrt{\sigma(t_c)}/2\pi \approx 0.1$ 
in Eq.~(\ref{eq:lc})
\footnote{
From Fig.~\ref{fig:X-x} we can read crossover wave number 
$\tilde{q} = q_c h^2 \approx 0.06$
and then 
$t_c = 1/q_c^2 \approx 280/h^{4}$ 
and 
$L_c = t_c^{1/2} \approx 17/h^{2}$. 
As a result 
$c_L = (1/2\pi) [C_2 \Delta_h L_c / \kappa^2]^{1/2} 
\approx 0.11$
where $C_2 \approx 0.056, \Delta_h=h^2/2$ 
and $\kappa=1$ corresponding to $J=1$.
}.
The growth rate of $\sigma(t)$ decreases quite quickly 
before $\sigma(t)$ reaches $(2\pi)^2$ (see Fig.~\ref{fig:sigma-t}). 
(In fact 
almost all of $|\theta_i|$ in our simulations are smaller than $\pi$ 
and the system does not {\it feel} that the potential is periodic.)
Therefore the early stage of the crossover, 
which has sufficiently long time range,  
hardly reflects the periodicity of the random potential 
and is similar to the crossover to the random manifold regime.

\acknowledgments

The present work is supported by 21st Century COE program 
``Topological Science and Technology'' 
and the Ministry of Education, Science, Sports and Culture, 
Grant-in-Aid for Young Scientists (A), 19740227, 2007.
A part of the computation in this work has been done 
using the facilities of the Supercomputer Center, 
Institute for Solid State Physics, University of Tokyo.



\end{document}